\documentclass[a4paper,11pt]{article}
\pdfoutput=1 % if your are submitting a pdflatex (i.e. if you have
\usepackage{jheppub} % for details on the use of the package, please
\usepackage{mathrsfs}

\usepackage{graphicx}

\usepackage[usenames,dvipsnames]{color}
\usepackage{amsmath}
\usepackage{bbm}
\usepackage{amsfonts}
\usepackage{amssymb}
\usepackage{latexsym}
\usepackage{graphicx}
\usepackage[english]{babel}
\usepackage{multirow}
\usepackage{float}
\usepackage{url}
\usepackage{slashed}
\usepackage{xcolor}

\newcommand{\be}{\begin{equation}}
\newcommand{\ee}{\end{equation}}
\newcommand{\ba}{\begin{array}}
\newcommand{\ea}{\end{array}}
\newcommand{\bea}{\begin{eqnarray}}
\newcommand{\eea}{\end{eqnarray}}

\usepackage{ulem,fancyvrb}
\usepackage{xcolor}

\title{CDF W mass anomaly from a dark sector with a Stueckelberg-Higgs portal}

\author[a]{Mingxuan Du}
\author[a,b]{Zuowei Liu}
\author[c]{Pran Nath}

\affiliation[a]{Department of Physics, Nanjing University, Nanjing 210093, China}
\affiliation[b]{CAS Center for Excellence in Particle Physics, Beijing 100049, China}
\affiliation[c]{Department of Physics, Northeastern University, Boston, MA 02115-5000, USA}

\emailAdd{mingxuandu@smail.nju.edu.cn}
\emailAdd{zuoweiliu@nju.edu.cn}
\emailAdd{p.nath@northeastern.edu}

\abstract{
We propose an explanation to the new W mass measurement recently reported by the CDF collaboration, which is larger than the standard model expectation by about 7 standard deviations. To alleviate the  tensions that are imposed on the electroweak sector by the new W mass measurement, we carry out an analysis in the Stueckelberg extended standard model where a new neutral gauge boson appears which mixes with the two neutral gauge bosons in the electroweak sector both via the Stueckelberg mass terms and via the gauge invariant Stueckelberg-Higgs portal interaction and spoils the custodial symmetry at the tree level so that the simple relation between the W boson mass and the Z boson mass does not hold. We find that such an extension increases the W boson mass if the new gauge boson mass is larger than the Z boson mass. We further show that there exists a significant part of the parameter space in the extended model which includes the CDF mass anomaly and is consistent with the various observables at the Z pole and consistent with the ATLAS dilepton limits. The Stueckelberg $Z'_{\rm St}$ boson, which resolves the CDF W mass anomaly, should be searchable in future LHC experiments. 

}

\begin{document}
\maketitle

\section{Introduction}

Recently, the $W$ boson mass has been accurately measured by the CDF II detector with 8.8 fb$^{-1}$ data accumulated at the Fermilab Tevatron collider with $\sqrt{s}=1.96$ TeV \cite{CDF:2022hxs}. 
The new $W$ mass measurement is 
\be
M_W^{\rm CDF} 
= 80,433.5 
\pm 6.4 \, ({\rm stat})     
\pm 6.9 \, ({\rm syst}) \, {\rm MeV} 
=  
80,433.5 
\pm 9.4 \, {\rm MeV}.
\ee
The standard model (SM) expectation for the W mass is  \cite{CDF:2022hxs} 
\begin{equation}
M_W^{\rm SM} = 80,357 \pm 6 \, {\rm MeV}.
\end{equation}
Thus the new CDF measurement on the W mass is larger than the SM expectation by the amount 
\be
\Delta M_W = + 76 \, {\rm MeV}, 
\ee
which corresponds to about 7 standard deviations, if the uncertainties are added in quadrature \cite{CDF:2022hxs}. 

Indeed early on a precise measurement of the $W^{\pm}$ and $Z$ masses was viewed 
as an important testing ground for the standard model~\cite{Marciano:1983wwa}. 
Thus in  the SM, the W mass and the Z mass are related via (at tree level, or in the on-shell scheme)
\begin{equation}
    M_W = M_Z \cos \theta_W, 
\end{equation}
where $M_Z$ is the Z boson mass, 
and $\theta_W$ is  {the} weak mixing angle. 
The value of $M_Z$ 
has been measured with great precision at LEP \cite{ParticleDataGroup:2020ssz}
\be
M_Z = 91.1876 \pm 0.0021 \, {\rm GeV}. 
\label{eq:ZMass}
\ee 
The weak mixing angle $\theta_W$ is also well measured via various Z pole observables by LEP \cite{ALEPH:2005ab}.
In the standard model where the gauge symmetry is broken by  {vacuum expectation value (VEV)} of the
Higgs doublet field, the  {so-called} custodial symmetry holds where 
$\rho\equiv \frac{M_W^2}{M_Z^2 \cos^2\theta_ {W}}=1$ holds at the  {tree} level and
$\rho$ is also very close to 1 including radiative corrections.
Thus, the new CDF W mass measurement with more than 7 $\sigma$ deviation from
the standard model hints at a  violation of the custodial symmetry. 
To explain the CDF W mass anomaly,   we carry out an analysis in the Stueckelberg extension
of the standard model (StSM)~\cite{Kors:2004dx,Feldman:2006ce,Kors:2005uz,Feldman:2006wb,Feldman:2007wj,Feldman:2009wv,Aboubrahim:2020lnr}
where the custodial symmetry does not hold at the tree level {(see also for related early work in string theory \cite{Antoniadis:2002cs,Ghilencea:2002da,Coriano:2005own} and the references therein.)}
The StSM has an
  extended gauge symmetry $SU(2)_L\times U(1)_Y\times U(1)_X$ where the
new gauge boson associated with $U(1)_X$ mixes with the hypercharge gauge boson of the SM via the Stueckelberg mass terms. As stated unlike  SM,  in StSM, the custodial symmetry no longer holds, because there are additional mass terms due to mass mixing with the new sector which
 leads to a  mass correction to the Z boson. We find that the mass correction to the Z boson is negative if the new  {$Z_{\rm St}'$} boson is heavier than the Z boson. We further show that the negative contribution to the Z boson mass can provide an explanation to the CDF W mass anomaly. The various Z pole observables that have been precisely measured at LEP provide very stringent constraints to beyond-the-standard-model modifications to the electroweak sector.  We carry out a detailed analysis by fitting various Z pole observables in the Stueckelberg extended model and show that the parameter space to interpret the CDF W mass anomaly can lead to a fit with a satisfactory total $\chi^2$ which is nearly as good as the original LEP fit \cite{ALEPH:2005ab}.

The rest of  the paper is organized as follows. 
In section \ref{sec:tree}, we illustrate the implications of the new CDF W mass  measurement to the electroweak sector and to the BSM models. 
In section \ref{sec:NewModel}, we present our model and provide an explanation to the W mass anomaly.  In section \ref{sec:zpole:fits}, we carry out a detailed analysis by fitting various Z pole observables in the new model. 
 {In section \ref{sec:results}, we present the results of our analysis and analyze constraints. 
In section \ref{sec:discuss}, we discuss some related constraints and implications.} 
In section \ref{sec:conclude}, we summarize our results. 
Some further details of the model are given in Appendix A.

\section{Implications of CDF W mass measurement}
\label{sec:tree}

We discuss the implications of the CDF W mass anomaly to the electroweak sector of the SM. We illustrate the effects using tree  {level} expressions in the SM. Although the expressions with the radiative corrections taken into account are more precise, they tend to be more complex and are often redefined with SM relations (which may or may not hold now). For the sake of clarity, we use the tree level expressions for the discussion in this section; the full analysis
  will result in small quantitative modifications which would not change the conclusions here. 

One commonly used quantity in the electroweak precision fitting is the Fermi constant, 
\begin{equation}
    G_F = \frac{g_2^2}{4\sqrt{2}M_W^2}, 
    \label{eq:fermi}
\end{equation}
where $g_2$ is the gauge coupling constant of the $SU(2)_L$ gauge group. Experimentally, the Fermi constant is determined precisely via muon lifetime measurement~\cite{ParticleDataGroup:2020ssz}
$
G_F = 1.1663787(6) \times 10^{-5} \, {\rm GeV}^{-2}. 
$
In the SM, the $W$ and $Z$ boson masses are 
\be
M_W = \frac{1}{2} g_2 v, \quad
M_Z = \frac{1}{2} \sqrt{g_2^2 + g_Y^2} v, 
\label{eq:WZmass}
\ee
where $g_Y$ is the gauge coupling constant of the $U(1)_Y$ gauge group, and $v$ is the  {VEV} % 
of the SM Higgs potential. Thus, a larger $M_W$ implies a larger $g_2$, if the relation given in Eq.\ \eqref{eq:fermi} is unchanged. 
The tree level $M_W$ expression also tells us that the Fermi constant $G_F$ is related to the Higgs
VEV by $G_F^{-\frac{1}{2}} =2^{\frac{1}{4}} v$ 
 which is then  unmodified by the new W mass measurement. Since the Z boson mass is precisely measured in LEP, the larger $g_2$ value then implies a smaller $g_Y$ if we wish to maintain the $M_Z$ relation of Eq.\ (\ref{eq:WZmass}). 
At the tree level in the SM, the weak mixing angle $\theta_W$ is given by 
\begin{equation}
    \sin^2\theta_W = \frac{g_Y^2}{g_2^2 + g_Y^2} 
    = 1 - \frac{M_W^2}{M_Z^2}. 
    \label{eq:weak}
\end{equation}
Using the central values of $M_W$ and $M_Z$, one finds that the deviation $\Delta M_W = + 76$ MeV leads to a shift 
\begin{equation}
    \Delta(\sin^2\theta_W) = -0.00147.
    \label{eq:s2w:change}
\end{equation}
Such a change on the weak mixing angle is severely constrained by the various Z pole observables; see the Z pole fit 
{labeled}  ``SM-CDF'' in Table \ref{tab:Table:2020} in which the above shift on the weak angle has been made while the rest of the SM expressions are unchanged as in the ``LEP'' fit. The total $\chi^2$ has increased from the LEP fit $\sim 17$ to $\sim 98$, which is in ballpark agreement with the 7 $\sigma$ deviation \cite{CDF:2022hxs}.

If we wish to hold the weak angle fixed, we require that 
$g_Y$ increase proportionately which implies  an increase in  the mass of the Z-boson 
above the current SM prediction by an amount 
\be
\Delta M_Z^{\text{exp}} =  \frac{\Delta M_W}{\cos\theta_W}.
\label{eq:delta1} 
\ee 
However, such an increase would be in severe conflict with the LEP measurement of the 
Z-boson mass given by Eq.\ \eqref{eq:ZMass}.
This problem is resolved in the Stueckelberg extension of the standard model where we show
that the Stueckelberg sector generates a negative contribution to the Z-boson mass which
cancels the contribution Eq.\ \eqref{eq:delta1} in a significant  
region of the parameter space restoring 
consistency with the CDF data.

\begin{table}[htbp]
\begin{centering}
\begin{tabular}{|c|c|c|c|c|c|c|c|}
\hline
  & & \multicolumn{2}{c|}{LEP}
  & \multicolumn{2}{c|}{SM-CDF}
  & \multicolumn{2}{c|}{St} \\  \hline
  & $O^{\rm exp}\pm \delta O$   
  & $O^{\rm th}$  & $\chi$  
  & $O^{\rm th}$  & $\chi$ 
  & $O^{\rm th}$ &  $\chi$ \\   \hline
%%%%%
% 
$\Gamma _Z$ [GeV] & 2.4955$\pm$0.0023 &  {2.4960} & -0.20 &  {2.5000} & -1.96 & 2.4999 & -1.92 \\      
$\sigma _{\text{had}}$ [nb] & 41.481$\pm$0.033 &  {41.470} & 0.34 & 41.465 & 0.50 & 41.471 & 0.30 \\      
$R_e$ & 20.804$\pm$0.05 & 20.752 & 1.05 & 20.778 & 0.53 & 20.751 & 1.05 \\      
$R_{\mu }$ & 20.784$\pm$0.034 & 20.752 & 0.95 & 20.778 & 0.18 & 20.751 & 0.96 \\      
$R_{\tau }$ & 20.764$\pm$0.045 & 20.799 & -0.77 & 20.825 & -1.35 & 20.798 & -0.76 \\      
$R_b$ & 0.21629$\pm$0.00066 & 0.21584 & 0.68 & 0.21578 & 0.77 & 0.21584 & 0.68 \\      
$R_c$ & 0.1721$\pm$0.003 & 0.1711 & 0.33 &  {0.1712} & 0.30 & 0.1711 & 0.33 \\      
$A_{\text{FB}}^{(0,e)}$ & 0.0145$\pm$0.0025 &  {0.0163} & -0.71 &  {0.0190} & -1.81 &  {0.0162} & -0.70 \\     
$A_{\text{FB}}^{(0,\mu )}$ & 0.0169$\pm$0.0013 &   {0.0163} & 0.48 &  {0.0190} & -1.64 &  {0.0162} & 0.51 \\   
$A_{\text{FB}}^{(0,\tau )}$ & 0.0188$\pm$0.0017 &   {0.0163} & 1.48 &  {0.0190} & -0.13 &  {0.0162} & 1.50 \\   
$A_{\text{FB}}^{(0,b)}$ & 0.0996$\pm$0.0016 &  {0.1033} & -2.29 &  {0.1118} & -7.61 &  {0.1032} & -2.22 \\     
$A_{\text{FB}}^{(0,c)}$ & 0.0707$\pm$0.0035 &  {0.0738} & -0.89 &  {0.0804} & -2.78 &  {0.0737} & -0.86 \\    
$A_{\text{FB}}^{(0,s)}$ & 0.0976$\pm$0.0114 &  {0.1034} & -0.51 &  {0.1119} & -1.25 &  {0.1033} & -0.50 \\    
$A_e$ & 0.15138$\pm$0.00216 & 0.14733 & 1.88 & 0.15928 & -3.66 & 0.14716 & 1.95 \\      
$A_{\mu }$ & 0.142$\pm$0.015 &  {0.147} & -0.36 &  {0.159}& -1.15 &  {0.147} & -0.34 \\      
$A_{\tau }$ & 0.136$\pm$0.015 &  {0.147} & -0.76 &  {0.159} & -1.55 &  {0.147} & -0.74 \\      
$A_b$ & 0.923$\pm$0.02 &  {0.935} & -0.58 &  {0.936} & -0.63 &  {0.935} & -0.58 \\      
$A_c$ & 0.67$\pm$0.027 &  {0.67} & 0.07 &  {0.67}  & -0.12 & {0.67}  & 0.08 \\      
$A_s$ & 0.895$\pm$0.091 &  {0.936} & -0.45 &  {0.937} & -0.46 &  {0.936} & -0.45 \\  \hline  
$\chi ^2$ &  &  & 17.3 &  & 97.8 &  & 20.9 \\     
\hline
\end{tabular}
\caption{Fits to 19 LEP Z pole observables \cite{Zyla:2020zbs} with three models: (1) LEP, (2) SM-CDF, and (3) St. 
The LEP fit is taken from \cite{ALEPH:2005ab}. 
The SM-CDF fit is the same as the LEP fit except that the weak mixing angle $\theta_W$ is modified as in Eq.\ \eqref{eq:s2w:change} due to the new CDF $M_W$  measurement \cite{CDF:2022hxs}. 
 The St Fit is described in the text; 
the parameters in the benchmark model here are:   
$M_1 = 725$ GeV,
$g_c$= $0.243$, 
$\bar{g}_Y = g_Y (1 + 0.047\%)$.}
\label{tab:Table:2020}
\end{centering}
\end{table}

\section{The Model}
\label{sec:NewModel}

In this section we consider an extended SM  model along with a hidden sector which possesses a $U(1)_X$ gauge symmetry. We consider 
a portal to the hidden sector consisting of two parts: one is the 
conventional Stueckelberg portal where there is mass mixing between
the hypercharge field $B_\mu$ and the Stueckelberg field $C_\mu$.
In addition we consider a second Stueckelberg-Higgs portal where
a $SU(2)_L\times U(1)_Y$ gauge covariant derivative of the Higgs doublet
couples to $C_\mu$. 
 Thus we write the total Lagrangian so that $\mathcal{L}= \mathcal{L}_{\text{SM}}+ \Delta {\cal L}_{\rm St}$ where
 \be
  {\Delta} \mathcal{L}_{\text{St}}= 
-\frac{C_{\mu \nu}^2}{4} 
-\frac{1}{2}\left(M_{1} \bar C_{\mu} +M_{2} B_{\mu} \right)^{2} +J^\mu_{\text{hid}} C_\mu 
+ \left(i\frac{g_c}{2} \Phi^\dagger D^\mu\Phi {\bar C_\mu} + h.c.\right)  
+  \frac{g_c^2}{4}  \Phi^\dagger \Phi
 \bar C_\mu^2. 
\label{extended-st}
\ee
Here $\Phi$ is the SM Higgs doublet, $D_{\mu}\Phi=\left[\partial_{\mu}- i \frac{{\bar{g}_{Y}}}{2}  B_{\mu}
- i \frac{{\bar{g}_2}}{2} {\tau^{a} A_{\mu}^{a}}\right]\Phi$ is the $SU(2)_L\times
U(1)_Y$ covariant derivative of $\Phi$, 
{$\tau_{a}$ is the Pauli matrix,} $\bar C_\mu=C_\mu+ \partial_\mu \sigma/M_1$
 is the combination of $C$ and the axion field $\sigma$
 which is gauge invariant  under the $U(1)_X$ gauge transformation, where
 $C_\mu$    couples to the hidden sector source $J^\mu_{\text{hid}}$,
$M_1$ and $M_2$ are the Stueckelberg 
mass parameters, and $g_c$ is a dimensionless parameter.
We note here that the first line of Eq.\ (\ref{extended-st}) is the standard
Stueckelberg portal of the visible sector to the hidden sector.
The second line of Eq.\ (\ref{extended-st}) is the new  Stueckelberg-Higgs
portal of the visible sector to the hidden sector. The coefficient $g_c^2/4$ 
of the $\bar C^2$ term is determined by the constraint that 
the neutral vector boson sector have a massless mode, i.e., the photon.\footnote{However, for the special case in which $M_2=0$, there is no constraint on the coefficient of the $\bar C^2$ term from the existence of the massless mode in  the neutral vector boson sector, since the determinant of the mass matrix of Eq.\ (\ref{mass-matrix}) vanishes for any value of the coefficient of the $\bar C^2$ term in this case.}
One may combine the Stueckelberg-Higgs portal terms in an extended
covariant derivative of the Higgs field so that 
$\mathcal{D}_\mu \Phi= (D_\mu - i\frac{g_c}{2} \bar C_\mu)\Phi$.

Here we also note that  the Lagrangian of Eq.\ (\ref{extended-st}) 
 is not invariant under $U(1)_Y$ unless either $M_2=0$ or $g_c=0$
 because invariance under hypercharge transformations 
 of $(M_1C_\mu + M_2 B_\mu +\partial_\mu \sigma)$
 and $(C_\mu+ \partial_\mu \sigma/M_1)$ cannot be simultaneously achieved.
 The case $g_c=0$ leads to the standard Stueckelberg model \cite{Kors:2004dx,Feldman:2006ce} while the case we consider in the analysis here is $M_2=0$.
 However, for the sake of generality we will keep both $M_2$ and  {$g_c$} %
 before we discuss the model  $M_2=0$  in further detail later.
%}

Note that we have used $\bar{g}_2$ and $\bar{g}_Y$ in Eq.\ (\ref{extended-st}),
because the gauge couplings in the new model do not necessarily take the same values as in SM.
Thus after spontaneous breaking of the electroweak symmetry and in the
unitary gauge the neutral gauge boson mass matrix 
in the basis  $V= (C, B, A^3)$  is given by
\begin{equation}
M^{2}=\left(\begin{array}{ccc}
M_{1}^{2}+\frac{v^{2}}{4} g_{c}^{2} & {M_{1}^2 \epsilon} +\frac{v^{2}}{4} g_{c} \bar{g}_{Y} & -\frac{v^{2}}{4} g_{c} \bar{g}_{2} \\
{M_{1}^2 \epsilon} +\frac{v^{2}}{4} g_{c} \bar{g}_{Y} & {M_{1}^2 \epsilon^2}+\frac{v^{2}}{4} \bar{g}_{Y}^{2} & -\frac{v^{2}}{4} \bar{g}_{Y} \bar{g}_{2} \\
-\frac{v^{2}}{4} g_{c} \bar{g}_{2} & -\frac{v^{2}}{4} \bar{g}_{Y} \bar{g}_{2} & \frac{v^{2}}{4} \bar{g}_{2}^{2}
\label{mass-matrix}
\end{array}\right), 
\end{equation}
where we have defined $\epsilon\equiv M_2/M_1$.
We note in passing that a restricted form of the 
mass matrix of Eq.\ (\ref{mass-matrix})
has been discussed previously \cite{Alguero:2022est} in an ad hoc fashion since no mechanism
was proposed %
for generation of the mass matrix.
The mass matrix has a vanishing determinant which ensures the existence of a massless photon mode. 
The mass matrix can be diagonalized via 
 {an orthogonal transformation ${\cal O}$  
such that $E_i={\cal O}_{ji} V_{j}$ and}  
${\cal O}^T M^2 {\cal O} = 
{\rm diag} (M^2_{Z'}, M^2_{Z}, 0)$.
The mass eigenstates $E_i = (Z', Z,  A_\gamma)$ are the new $Z'$, the $Z$, and the photon.
The interaction Lagrangian between 
the massive neutral gauge bosons
and the SM fermion $f$ is given by 
\be
{\cal L}_{NC} =
\bar f \gamma^\mu \left[ 
 ( v_{f} -\gamma_5 a_{f})Z_\mu 
+ (v'_{f} -\gamma_5 a'_{f})Z'_\mu 
 \right ] f 
 + e \bar f \gamma^\mu Q_f  A_\mu f, 
\ee
where 
\bea
a_f &=& \sqrt{\rho_f}(\bar{g}_Y {\cal O}_{22} - 
\bar{g}_2 {\cal O}_{32})T^{3 }_f/2 , \\
v_f &=& a_f
- \sqrt{\rho_f} \kappa_f \bar{g}_Y {\cal O}_{22} Q_f, \\
a'_f &=& (\bar{g}_Y {\cal O}_{21} - 
\bar{g}_2 {\cal O}_{31})T^{3 }_f/2, \\
v'_f &=& a'_f - \bar{g}_Y {\cal O}_{21}Q_f. 
\eea
Here $Q_f$ and $T_f^3$ are the electric charge 
and SU(2)$_L$ quantum number of the fermion $f$. 
For the Z couplings, we have incorporated the $\rho_f$ and $\kappa_f$ factors that contain radiative corrections from propagator self energies and flavor specific vertex corrections and 
are given in Ref.\ \cite{Erler:2004nh, ALEPH:2005ab}. 
{For the tree level expressions, $\rho_f \to 1$ and $\kappa_f \to 1$.} 

In this new model
the $W$ mass is predicted as
$M_W = \bar{g}_2 v /2$.
To explain the larger W mass measured by CDF, we increase the $SU(2)_L$ gauge coupling 
\begin{equation}
    \bar{g}_2 = g_2 
    \left( 1 + \frac{\Delta M_W}{M_W} \right). 
    \label{eq:g2bar}
\end{equation}
Since $M_W=\bar{g}_2 v/2$ appears  {in} the $(3,3)$ element in the neutral boson mass matrix, it 
{tends}
to increase the eigenvalue $M_Z$.
{In the Stueckelberg extended standard model}
 we compensate the increase of the $Z$ mass due to 
the new CDF measurement on the $W$ mass
with a  negative $Z$ mass correction 
from the Stueckelberg mass shift of the Z boson  arising from mixing of the
neutral gauge  {bosons} of the standard model with  {the} Stueckelberg gauge boson.

{As discussed earlier
 there are two model classes}: (1) $g_c=0$, and (2) $M_2=0$. In the former case, the model is the StSM~\cite{Kors:2004dx,Feldman:2006ce} 
and here we find that the StSM has a total $\chi^2$ for the Z pole fitting which is not better than the ``SM-CDF'' fit. Thus, in the present analysis we  
  discuss  (2) when $M_2=0$.  In this case the Stueckelberg mass correction is given by
    \begin{align}
\Delta M_Z^{\rm th} \simeq -\frac{1}{2} M_Z 
\frac{M_W^2}{M_1^2-M_Z^2} \left(\frac{g_c}{g_2} \right)^2.  
\label{eq:delta:th} 
  \end{align}
  The Stueckelberg mass correction to the Z-boson is positive for $M_1<M_Z$ 
and negative for $M_1>M_Z$.  The branch favored by the CDF W mass measurement is
$M_1>M_Z$ and here one can achieve a cancellation, i.e., $\Delta M^{\rm exp}_{Z}+
\Delta M^{\rm th}_{Z} \simeq 0$.   

\section{Fitting Z pole observables}
\label{sec:zpole:fits}

To ensure that the modifications in the neutral sector respect the electroweak data, we perform a global analysis by fitting the precision electroweak LEP data~\cite{ALEPH:2005ab} {similar to the analysis done in Ref.\ \cite{Feldman:2006ce}}.
There are {two} 
new parameters in the model: 
$M_1$
and $g_c$. Unlike SM, the relation given in Eq.\ \eqref{eq:weak} no longer holds in the new model. Thus, we also let $\bar{g}_Y$ vary in our analysis.
In the global fits
we constrain 
the model by 
fitting the 
$Z$ mass \cite{ParticleDataGroup:2020ssz} and the other
19 $Z$ pole quantities
\cite{Zyla:2020zbs},
including the total $Z$ decay width $\Gamma_Z$,
the hadronic pole cross-section  {$\sigma_{\rm had}$},
ratios of branching ratios $R$,
the various forward and backward asymmetries {$A_f$ and $A_{\rm FB}$}.

The partial decay width of the $Z$ into a pair of fermions is given by 
\cite{Baur:2001ze,Erler:2004nh,ALEPH:2005ab,Feldman:2006wb}
\be
\Gamma_{ff}  = 
N_f^c  \mathcal{R}_f \frac{M_Z}{12\pi}  \sqrt {1-4\mu _f^2 } 
\times 
\Bigg  [ |v_f|^2 (1 + 2\mu _f^2 ) + |a_f|^2 (1 -4\mu _f^2 )\Bigg], 
\ee
where $N_f^c =(1,3)$ for leptons and quarks,
$\mu _f = m_f  / M_Z$, 
and $\mathcal{R}_f$ are the radiative corrections, which are given by 
\bea
 \mathcal{R}_f  &=& \left(1 + \delta _f^{\rm QED} \right)\left(1 +\frac{N_f^c-1}{2}\delta _f^{\rm QCD} \right),\\
 \delta _{f}^{\rm QED}  &=&\frac{3\alpha}{4\pi}Q_f^2 , \\
 \delta _{f}^{\rm QCD}  &=& \frac{{\alpha_s }}{\pi } + a_1 \left[ {\frac{{\alpha _s }}{\pi }} \right]^2
 + a_2 \left[ {\frac{{\alpha _s }}{\pi }} \right]^3  - Q_f^2
\frac{{\alpha \alpha _s   }}{{4\pi ^2 }}, 
\eea
where $a_1 = 1.409$ and $a_2=-12.77$. 
Here  $\alpha $ and  $\alpha _s $ 
are taken at the $M_Z$ scale. 
Using the partial width, one can then compute the 
total $Z$ decay width $\Gamma_Z$, and 
\be
\sigma_{\rm had} \equiv \frac{12 \pi}{M_Z^2}
\frac{\Gamma_{ee} }{\Gamma_Z}
\sum_{q\neq t} \frac{\Gamma_{qq}}{\Gamma_Z},
\quad
R_{\ell} \equiv \frac{\Gamma_{\rm had}}{\Gamma_{\ell \ell}},
\quad
R_q \equiv \frac{\Gamma_{qq}}{\Gamma_{\rm had}}. 
\ee
The asymmetry parameters $A_f$ 
and $A_{\rm FB}^{(0,f)}$   {are} defined in terms of 
 the fermion couplings to the $Z$ \cite{ALEPH:2005ab}
\be
A_f = \frac{2 v_f a_f}{v_f^2 + a_f^2}, \quad
A_{\rm FB}^{(0,f)} = \frac{3}{4} A_e A_f. 
\ee
We compute $\chi_i$ for each quantity via 
\be
 \chi_i= \frac{O^{\rm exp}_i - O^{\rm th}_i}{\delta O_i},  
\ee
where $O^{\rm exp}_i$, $\delta O_i$ and $O^{\rm th}_i$ are 
the central value and uncertainty of LEP results, 
and the new model expectation in that order 
for the Z-pole quantities
shown in the first column of Table \ref{tab:Table:2020}. 
The total $\chi^2$ is computed via $\chi^2 = \sum_i \chi_i^2$.

\section{Results}
\label{sec:results}

\begin{figure}[thbp]
\begin{centering} 
\includegraphics[width=0.5 \textwidth]{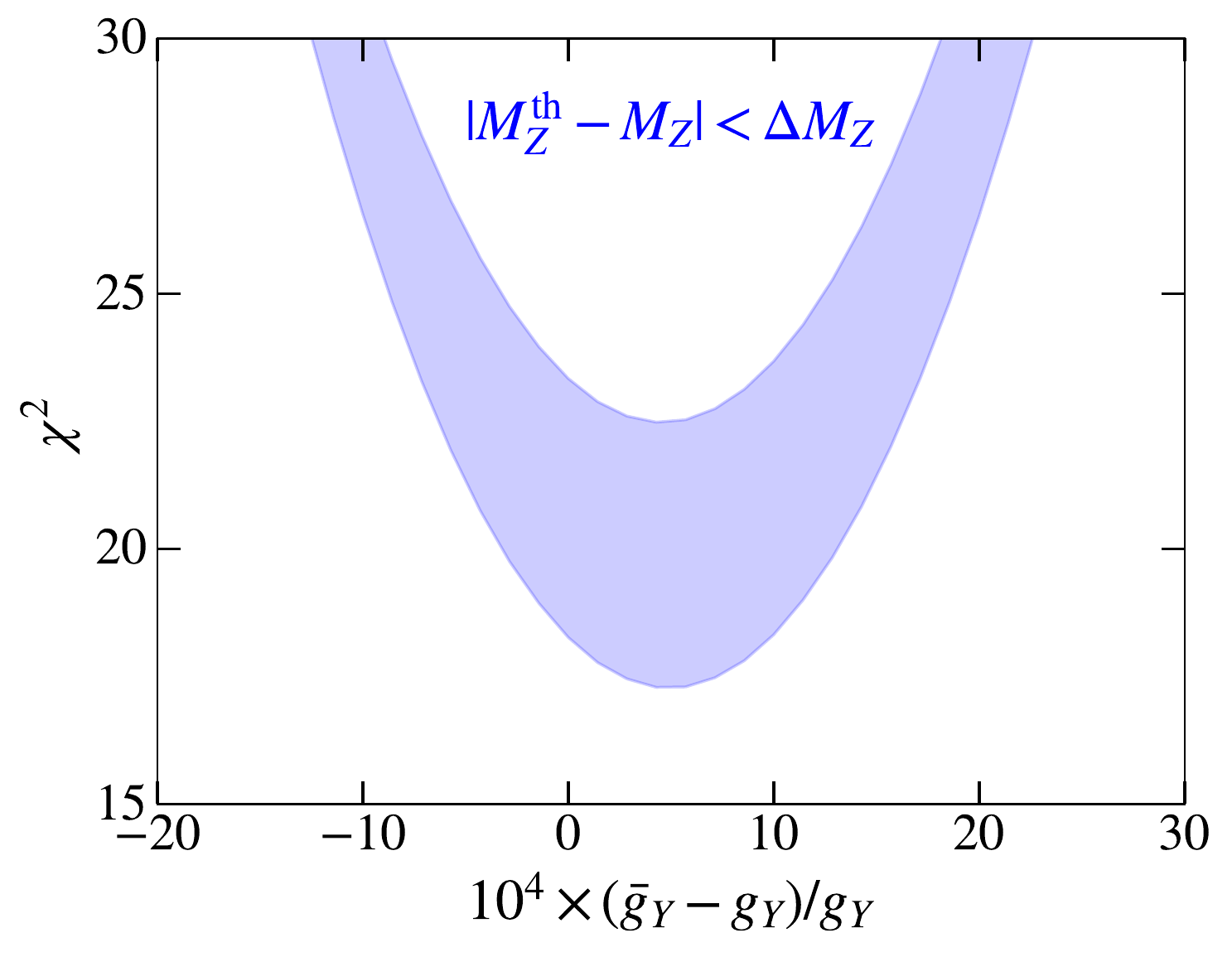}
\caption{Total $\chi^2$ of 19 Z pole observables versus $\bar{g}_Y$ for the model points that reproduce the Z mass consistent with the experimental value within an error bar of 2.1 MeV \cite{ParticleDataGroup:2020ssz}.} 
\label{fig:gYbest}
\end{centering}
\end{figure}

We carry out Monte Carlo scans in the parameter space spanned by $M_1$, $g_c$, and $\bar{g}_Y$. In the calculation, we first fix the parameters $g_2$, $g_Y$, and $v$ to the SM values such that the canonical W and Z masses  {and $G_F$} can be obtained using Eqs.\ (\ref{eq:fermi}-\ref{eq:WZmass}). We then shift % 
 {$g_2$} to be $\bar{g}_2$ according to Eq.\ \eqref{eq:g2bar}.
For each model point, we require that the mass matrix leads to the canonical Z mass within the %
 {uncertainty} of $\sim 2$ MeV \cite{ParticleDataGroup:2020ssz}, %
{and} compute the total $\chi^2$ for the 19 Z pole observables as shown in Table \ref{tab:Table:2020}. We find that the total $\chi^2$ has a strong dependence on $\bar{g}_Y$, as shown in Fig.\ \ref{fig:gYbest}. The model points with the least $\chi^2$ are located in the region 
{where $\bar{g}_Y$} is about 0.047\% {larger than} the SM value. We show one benchmark model point in Table \ref{tab:Table:2020} with 
 {$\chi^2 \simeq 21$.}

\begin{figure}[htbp]
\begin{centering}
\includegraphics[width=0.5 \textwidth]{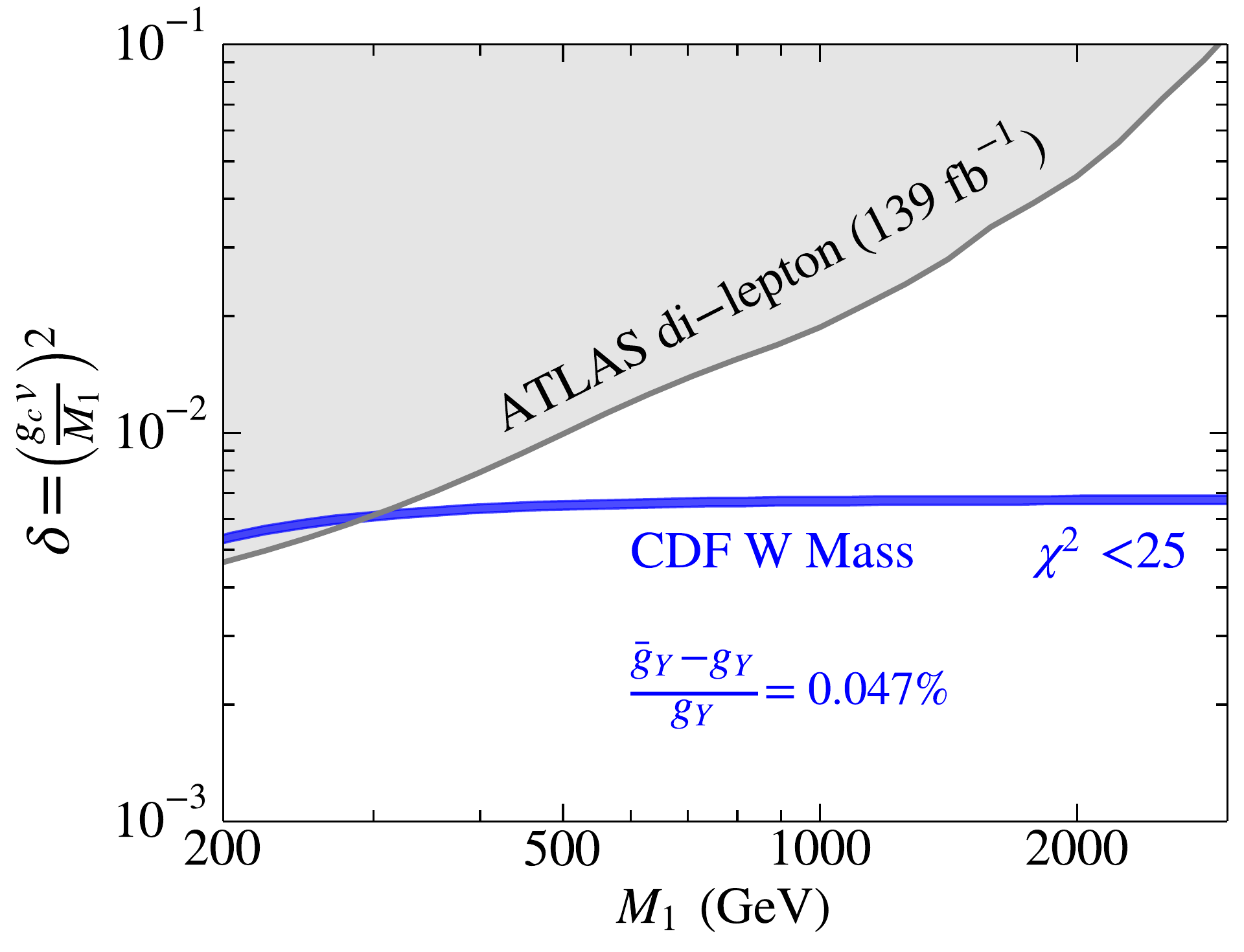}
\caption{ {The favored  {region} % 
(blue) in the parameter space 
spanned by $M_{1}$ and $\delta = (g_c v/M_1)^2$ 
for the CDF $W$ mass measurement \cite{CDF:2022hxs} 
and $\chi^2<25$ for fitting the 19 Z pole observables, 
where we fix the $\bar{g}_Y =g_Y (1+0.047\%)$.
The constraints 
from di-lepton searches at ATLAS \cite{ATLAS:2019vcr} 
are also shown.}}
\label{fig:CCDF-scanD}
\end{centering}
\end{figure}

Thus we fix $\bar{g}_Y$ to be the value corresponding to the least $\chi^2$, which is 
$\bar{g}_Y =g_Y (1+0.047\%)$, 
and further carry out Monte Carlo scans in the 2D parameter space spanned by $M_1$ and $g_c$. The results are given in Fig.\ \ref{fig:CCDF-scanD} where 
the favored region for the CDF $W$ mass measurement \cite{CDF:2022hxs} 
and with $\chi^2 < 25$ for fitting the Z-pole quantities and Z mass in the parameter space spanned by $M_{1}$ and
 {$\delta \equiv (g_c v/M_1)^2$} are shown. 
It is found that the blue region can be well approximated by the following relation 
\be
\frac{1}{8} \left( \frac{g_c v}{M_1}\right)^2 \simeq \frac{ \Delta M_W}{M_W}, 
\ee
which is consistent with Eq.\ \eqref{eq:delta1} and Eq.\ \eqref{eq:delta:th}.
Di-lepton final states which are from the heavy $Z'$ boson at the LHC
can be searched for by
reconstructing their invariant mass.
The excluded region by
the di-lepton high mass resonance searches at ATLAS
\cite{ATLAS:2019vcr}
is shown 
as the shaded region 
in  Fig.\ \ref{fig:CCDF-scanD}.
Thus the parameter space consistent with both the CDF W mass measurement and the ATLAS di-lepton constraints is $M_1 \gtrsim 300$ GeV.

\section{Discussion}
\label{sec:discuss}

Further, we note that the increase in the mass of the W boson would 
lead to a larger width for the W-boson from its current value of 
$\Gamma_W=2.085$ GeV to a new value of $2.091$ GeV, 
which are consistent with the current uncertainty on the W width 
measurement, $\Gamma_W=2.085\pm 0.042$ GeV \cite{ParticleDataGroup:2020ssz}.
On the other hand, 
the  shift in the standard model electroweak correction to $g_\mu-2$ 
due to an  increased value of the W-boson is totally negligible, i.e., 
 $O(<1) \times 10^{-11}$ compared
to the standard model electroweak correction which is $a_{\mu}^{\rm EW} (\text{1-loop})
=194.8\times 10^{-11}$~\cite{Muong-2:2021ojo}. 
One of the predictions of the model is the possibility of observation of the $Z'$ boson
consistent with CDF W mass anomaly which could be searched for at the LHC.
The electroweak data constrains 
the particle physics models including $Z'$ physics; see e.g., 
\cite{Carena:2004xs,Langacker:2008yv,Kim:2011xv}. 
We note  in passing that in the usual  standard  Stueckelberg extension of the
standard model ~\cite{Kors:2004dx,Feldman:2007wj,Cheung:2007ut,Aboubrahim:2021ohe,Li:2021kso} which here is the case $g_c=0$, inclusion of matter in the $U(1)_X$ hidden sector will make the hidden sector matter  milli-charged.
However, in the model we discussed here where $M_2=0$, the hidden
sector matter does not develop a millicharge since the field $C_\mu$ 
has no photonic content. 
 {This is due to the vanishing ${\cal O}_{13}$ element in the rotation matrix in the $M_2=0$ case, as shown in appendix \ref{sec:mass:rotation}.} 

\section{Conclusion}
\label{sec:conclude}

In summary, it is shown that there exists a significant parameter space in the 
Stueckelberg extension of the standard model which can simultaneously
explain the CDF $W$ mass measurement,
fit the LEP electroweak quantities well 
(with $\chi^2$ as low as $\chi^2 \simeq 20$) 
and avoid the ATLAS di-lepton search constraint~\cite{ATLAS:2019vcr}.
The model is testable at the LHC because of the predicted existence of the
Stueckelberg  {$Z_{\rm St}'$} boson.

Finally we note that several works have appeared since the new CDF W mass measurement became public 
\cite{Zhu:2022tpr,
Fan:2022dck,
Lu:2022bgw,  %%%%
Strumia:2022qkt,
Athron:2022qpo,
Yang:2022gvz,
deBlas:2022hdk,
Tang:2022pxh,
Du:2022pbp,
Campagnari:2022vzx,
Cacciapaglia:2022xih,
Yuan:2022cpw, 
Blennow:2022yfm,
Sakurai:2022hwh,
Fan:2022yly,
Liu:2022jdq, %%%% 
Zhu:2022scj,
Arias-Aragon:2022ats,
Paul:2022dds,
Babu:2022pdn,
DiLuzio:2022xns,
Bagnaschi:2022whn,
Heckman:2022the,
Lee:2022nqz,
Cheng:2022jyi,
Bahl:2022xzi,
Song:2022xts,
Asadi:2022xiy,
Athron:2022isz,
Heo:2022dey,
Crivellin:2022fdf,
Endo:2022kiw,
Du:2022brr,
Cheung:2022zsb,
DiLuzio:2022ziu,
Balkin:2022glu,
Biekotter:2022abc,
Krasnikov:2022xsi,
Zheng:2022irz,
Ahn:2022xeq,
Kawamura:2022uft,
Peli:2022ybi,
Ghoshal:2022vzo,
Perez:2022uil,
Nagao:2022oin,
Kanemura:2022ahw,
Mondal:2022xdy,
Zhang:2022nnh,
Cirigliano:2022qdm,
Borah:2022obi,
Chowdhury:2022moc,
Arcadi:2022dmt,
Han:2022juu,
Gu:2022htv}.

\section{Acknowledgement}

The research of 
MD and ZL was supported in part by the National Natural Science Foundation of China under Grant No.\ 11775109, 
while the research of 
PN was supported in part by the NSF Grant PHY-1913328.

\appendix 

\section{The rotation matrix ${\cal O}$}
\label{sec:mass:rotation}

The orthogonal matrix ${\cal O}$ that diagonalizes the mass matrix of Eq.\ (\ref{mass-matrix}) when $M_2=0$ can be parameterized by two angles as follows 
\begin{equation}
{\cal O} = \left(\begin{array}{ccc}
\cos\phi & \sin\phi & {0} \\
\sin\theta \cos\phi  & -\sin\theta \cos\phi & \cos\theta \\
-\cos\theta \sin \phi  & \cos\theta \cos\phi & \sin\theta
\end{array}\right), 
\end{equation}
where 
\begin{equation}
    \cos\theta = \frac{\bar{g}_2}{\sqrt{\bar{g}_2^2 + \bar{g}_Y^2}}, \quad
    \sin\theta = \frac{\bar{g}_Y}{\sqrt{\bar{g}_2^2 + \bar{g}_Y^2}}, 
\end{equation}
and 
\begin{equation}
    \tan(2 \phi) = \frac{2 M_c M_{\tilde{Z}}}{M_1^2 + M_c^2 -M_{\tilde{Z}}^2}, 
\end{equation}
with     
\bea 
M_c = \frac{1}{2} g_c v, \quad
M_{\tilde{Z}} = \frac{1}{2} \sqrt{\bar{g}_2^2 + \bar{g}_Y^2} v. 
\eea
The ratio of ${\cal O}_{22}$ to ${\cal O}_{32}$ is given by
\begin{equation}
\frac{{\cal O}_{22}}{{\cal O}_{32}} = -\tan\theta 
= - \frac{\bar{g}_Y}{\bar{g}_2}. 
\end{equation}
The angle $\theta$ is effectively the weak mixing angle measured by the various forward-backward symmetrices at LEP I.  
One may compare the result above with that in  SM where the Z boson is 
given by
$Z = \cos\theta_W A^3 - \sin\theta_W B$, and 
where $\tan\theta_W = \frac{{g}_Y}{{g}_2}$. 
Finally we note that since ${\cal O}_{13}=0$, $C_\mu$ has no component
that connects it to the photon, and thus the hidden sector dark matter if it
exists would not receive a milli-charge as happens in the  conventional
StSM model.

\end{document}